\documentclass[prd,tightenlines,nofootinbib,showpacs,preprintnumbers,superscriptaddress, twocolumn]{revtex4}
\usepackage{amsfonts,amsmath,amssymb,amsthm,bbm,hyperref}
\usepackage{graphicx}
\usepackage{color}
\usepackage[T1]{fontenc}

\newcommand{\be}{\begin{equation}}
\newcommand{\ee}{\end{equation}}
\newcommand{\beq}{\begin{eqnarray}}
\newcommand{\eeq}{\end{eqnarray}}

\begin{document}

\title{Inter-universal entanglement in a cyclic multiverse}
\author{Salvador Robles-P\'{e}rez}
\affiliation{Instituto de  F\'{\i}sica Fundamental, CSIC, Serrano 121, 28006
Madrid, Spain}
\affiliation{Estaci\'{o}n Ecol\'{o}gica de Biocosmolog\'{\i}a, Pedro de Alvarado, 14, 06411 Medell\'{\i}n, Spain.}
\author{Adam Balcerzak}
\affiliation{Institute of Physics, University of Szczecin, Wielkopolska 15, 70-451 Szczecin, Poland}
\affiliation{Copernicus Center for Interdisciplinary Studies, S{\l}awkowska 17, 31-016 Krak{\'o}w, Poland}
\author{Mariusz P. D\k{a}browski}
\affiliation{Institute of Physics, University of Szczecin, Wielkopolska 15, 70-451 Szczecin, Poland}
\affiliation{National Centre for Nuclear Research, Andrzeja So{\l}tana 7, 05-400 Otwock, Poland}
\affiliation{Copernicus Center for Interdisciplinary Studies, S{\l}awkowska 17, 31-016 Krak{\'o}w, Poland}
\author{Manuel Kr{\"a}mer}
\affiliation{Institute of Physics, University of Szczecin, Wielkopolska 15, 70-451 Szczecin, Poland}

\date{\today}

\begin{abstract}
We study scenarios of parallel cyclic multiverses which allow for a different evolution of the physical constants, while having the same geometry. These universes are classically disconnected, but quantum-mechanically entangled. Applying the thermodynamics of entanglement, we calculate the temperature and the entropy of entanglement. It emerges that the entropy of entanglement is large at big bang and big crunch singularities of the parallel universes as well as at the maxima of the expansion of these universes. The latter seems to confirm earlier studies that quantum effects are strong at turning points of the evolution of the universe performed in the context of the timeless nature of the Wheeler-DeWitt equation and decoherence. On the other hand, the entropy of entanglement at big rip singularities is going to zero despite its presumably quantum nature. This may be an effect of total dissociation of the universe structures into infinitely separated patches violating the null energy condition. However, the temperature of entanglement is large/infinite at every classically singular point and at maximum expansion and seems to be a better measure of quantumness.
\end{abstract}

\maketitle


\section{Introduction}

The idea of parallel universes due to Everett \cite{Everett} and its more exotic extensions \cite{TegmarkSA,Hartle15} has been put into a more mathematical shape within the framework of the superstring landscape \cite{Susskind} (though not without doubts \cite{bankstop500}) and now is taken more and more seriously as a hypothesis testable by observations.         

One of the key points of a possible verifiability of such an idea is the fact that some classically disconnected regions of spacetime or universes can be quantum-mechanically entangled and this entanglement can have some influence on observational quantities in our universe or in each universe of the whole set known as the multiverse. In Ref. \cite{Laura09}, for example, it was suggested that the dark flow of matter in our universe -- as represented by an extra cosmic microwave background (CMB) temperature dipole -- could be due to the 
quantum-mechanical interference of our universe with the other universes of the multiverse. More effects, such as the suppression of the power spectrum at large angular scales, running of the spectral index, and a suppression of the $\sigma_8$ parameter have been suggested to result from having an extra contribution to an average Friedmann equation describing our universe due to quantum entanglement \cite{kinney16}. Possible deviations from the standard CMB perturbations spectrum in the context of landscape multiverse inflationary potentials have been studied recently as well \cite{Laura16a,Laura16b}. 

The idea of quantum entanglement is a well-established area of physics and enters into such disciplines like quantum information, quantum cryptography, quantum-dense coding, computational algorithms, quantum teleportation and many others \cite{horodeckiRMP,nakagawa16,Sengupta16}. It has also been considered in the context of cosmology and astrophysics in numerous papers \cite{Lee15,Baskal16,nishioka,Casini}. Very interesting features of the entanglement of particle physics processes have been found \cite{Latorre}, including the entanglement of four photons \cite{Hiesmayr}. 

The most natural framework for investigations of entanglement is quantum cosmology \cite{kieferbook}. However, while one of the main formulations of canonical quantum gravity is based on the Wheeler-DeWitt equation, the best formulation which can be used for calculations with regard to the quantum entanglement problem is the third quantization picture in which creation and annihilation operators for universes are postulated \cite{Strominger,pimentel01,kiefer93}. This formulation was used to discuss the problem of entanglement in a quantum-cosmological picture \cite{salva2010,salva2014,salva2015}. 

Besides, in the third quantization picture, one is able to describe the quantum-mechanical scheme for the birth of baby universes \cite{Strominger}. An interesting problem is how one gets new universes as separate entities (``the separate universe problem'') within the framework of the classical and quantum picture \cite{CDL,CH,Kopp,Carr14,Dai15,Quintin16,Oshita}. 

In this paper, we will be interested in extending the discussion of Ref.~\cite{Marosek2016} of classical cyclic universes or multiverses originally based on the idea of Tolman \cite{tolman,mnras95} and on the idea of varying constants \cite{regular2013} onto the quantum-mechanical picture of entanglement, and relate it to the problem of decoherence and the arrow of time in cosmology \cite{packets,timearrow,PRD2006}. As a starting point, the quantum-cosmological picture will be applied \cite{vilenkintunnel,HH,Atkatz}. A previous point related to that was that some strong quantum effects are possible 
 at the turning point of the evolution of the universe \cite{packets,oscill96,tunnel95} -- later the scenario was dubbed as a simple harmonic universe (SHU) in Refs.~\cite{Mithani12,Mithani14}. 

The paper is organized as follows: In section \ref{classical} we present the classical picture of cyclic universes evolving parallelly in the multiverse. In section \ref{quantum} we describe the formalism of quantum entanglement in the context of the multiverse and in section \ref{thermodynamics} we calculate the temperature and entropy of entanglement for the cosmological models under study. In section \ref{conclusions} we give our conclusions. 

\section{Classical cyclic multiverses}
\label{classical}

In Ref.~\cite{regular2013}, the theories of varying physical constants (gravitational constant $G$ and speed of light $c$) have been applied to remove, soften or change the nature of various singularities in cosmology. The mechanism is based on alternative gravity theories which allow for some extra fields in the gravitational sector and can be responsible for a different evolution of the universe. One of the well-known examples of such an approach is the ekpyrotic/cyclic model in which the evolution of the universe is non-singular in a four-dimensional spacetime (a brane) due to some special coupling of a scalar field representing the gravitational coupling to the Lagrangian \cite{Khoury2004,Turok2005}. Another example is the influence of quantum effects represented by higher-order corrections to the action \cite{Houndjo}. 

\begin{figure}
\centering
\includegraphics[width=0.5\textwidth]{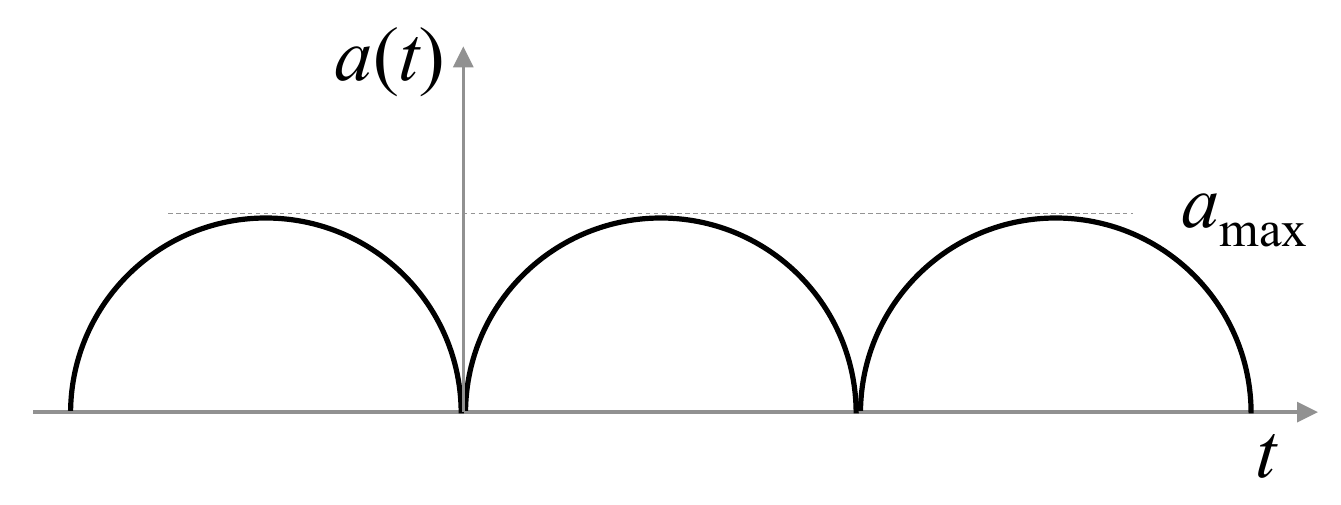}
\caption{Scale factor for the cyclic multiverse (sinusoidal pulse).}
\label{fig1}
\end{figure}

Here, we construct a toy model in which we allow for the variability of the gravitational constant $G$ (and thus the gravitational coupling constant) within the framework of the simple Friedmann geometry, such that the Einstein-Friedmann equations generalize to \cite{BM}  
\begin{eqnarray}
\label{rho1} \rho(t) &=& \frac{3}{8\pi G(t)}
\left(\frac{\dot{a}^2}{a^2} + \frac{kc^2}{a^2}
\right)~,\\
\label{p1} p(t) &=& - \frac{c^2(t)}{8\pi G(t)} \left(2 \frac{\ddot{a}}{a} + \frac{\dot{a}^2}{a^2} + \frac{kc^2}{a^2} \right)~,
\end{eqnarray}
and the energy-momentum ``conservation law'' is modified to
\begin{equation}
\label{conser}
\dot{\rho}(t) + 3 \frac{\dot{a}}{a} \left(\rho(t) + \frac{p(t)}{c^2} \right) = - \rho(t) \frac{\dot{G}(t)}{G(t)}~.
\end{equation}
The classical behavior of cyclic models of the universe (with finite values of the mass density and pressure at the turning points) due to the dynamics of the gravitational constant with pulses starting from a big bang and terminating at a big crunch which then again becomes a big bang based on the equations (\ref{rho1})--(\ref{conser}) has been analysed in Ref.~\cite{Marosek2016}. These models assumed a special type of the scale factor, which we will refer to as ``sinusoidal pulse'' in the following (see Fig.~\ref{fig1}), given by
\be
\label{sinusoidal}
a(t) = a_0 \left| \sin\left(\pi\frac{t}{t_c} \right) \right| ,
\ee
where $a_0, t_c=$ const., and a varying gravitational constant given by 
\be
\label{VC01}
G(t) = \frac{G_0}{a^2(t)} .
\ee
Assuming a closed universe with a constant velocity of light $c$, the energy density is equal to
\be
\rho(t) = \frac{3}{8 \pi G_0 } \left[  \frac{\pi^2}{t_c^2} \left( a_0^2 - a^2 \right)+ c^2 \right] > 0 ,
\ee
where $a\in(0,a_0)$. The Friedmann equation reads as
\be
\label{FE01}
H^2 \equiv \frac{1}{a^2} \left( \frac{d a}{d t}\right)^2 = \frac{\pi^2}{t_c^2} \left( \frac{a_0^2}{a^2} - 1 \right)  .
\ee
Even though we have taken a positive curvature, $k=+\,1$, in (\ref{FE01}), this equation can be considered as equivalent to the evolution equation of an open anti-de Sitter universe, for which the Friedmann equation reads as
\be\label{FE02}
H^2 = - \Lambda + \frac{1}{a^2} ,
\ee
provided that we choose 
\be
\label{Lama0}
\Lambda \equiv \frac{\pi^2}{t_c^2} \hspace{0.2cm} {\rm and} \hspace{0.2cm} a_0=\frac{1}{\sqrt{\Lambda}}. 
\ee
Besides, the relation (\ref{VC01}) gives a timeless trajectory in configuration space
\be
G(a) = \frac{G_0}{a^2} 
\ee
for the two variables $(a, G)$ \cite{packets}. 

Following Ref.~\cite{Marosek2016} one is able to extend this cyclic model into at least two universes of the same geometry, but with a different evolution of the gravitational constants in each of them.

\begin{figure}
\centering
\includegraphics[width=0.5\textwidth]{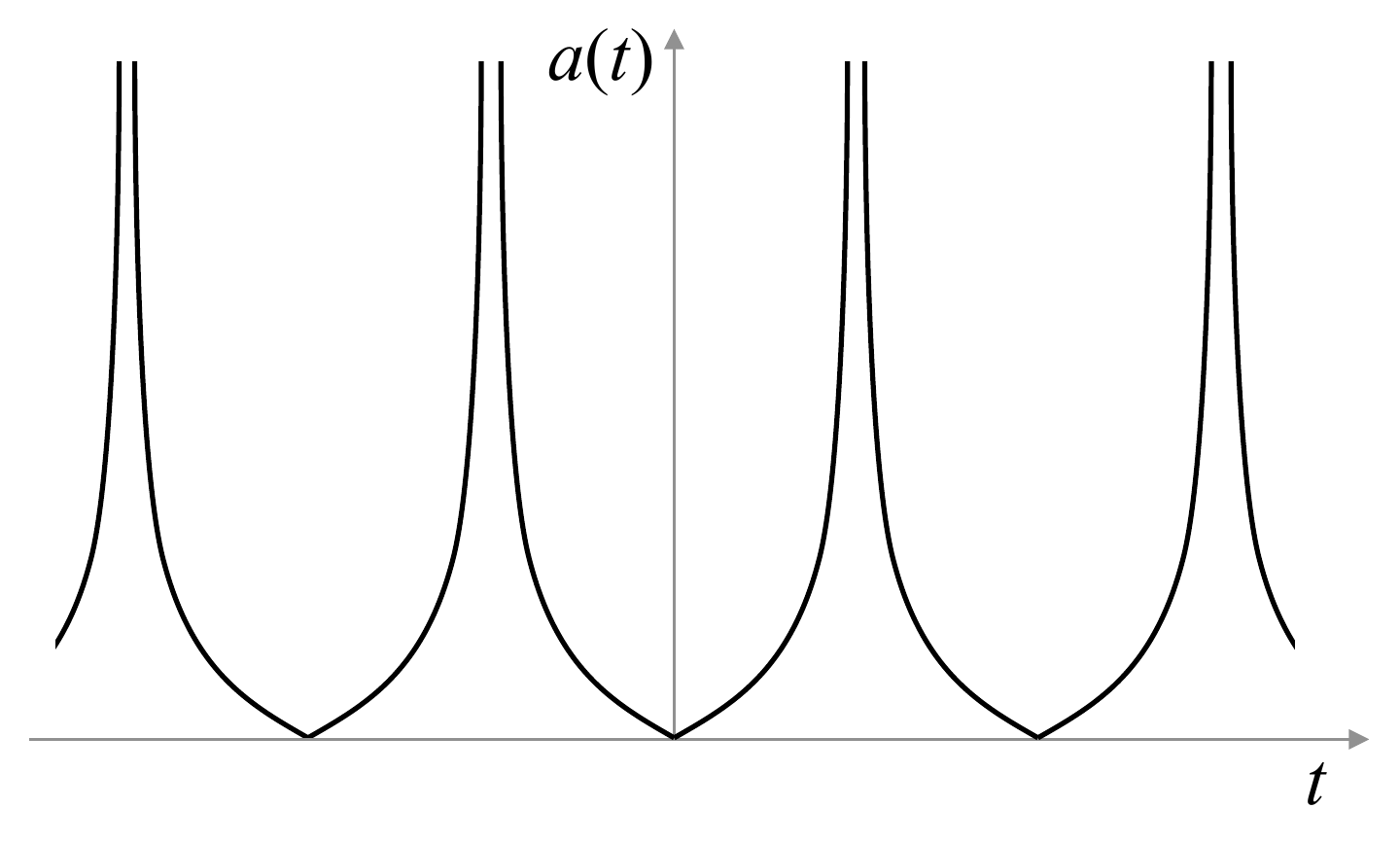}
\caption{Scale factor for the cyclic multiverse (tangential pulse).}
\label{fig2}
\end{figure}

Another example of a cyclic universe of Ref. \cite{Marosek2016} (with finite values of the mass density and pressure at the turning points) with pulses starting at a big bang and terminating at a big rip (see, Fig. \ref{fig2}), which then connects to a big bang, is possible when one chooses the scale factor to be
\be
\label{acyclic2}
a(t)= a_{0} \left| \tan \left( \pi \frac{t}{t_{s}} \right) \right|,
\ee
where $a_0, t_s=$ const., and the gravitational constant to vary as
\be
\label{Gcyclic2}
G \left( t \right) = \frac{4G_s}{\sin^2{\left( 2 \pi \frac{t}{t_s} \right)}}.
\ee
The timeless trajectory in configuration space for (\ref{acyclic2}) and (\ref{Gcyclic2}) is given by 
\be
\label{tantrajectory}
G(a) = \frac{G_s}{a_0^2} \,\frac{(a^2 + a_0^2)^2}{a^2} ,
\ee
which shows that at both the big bang $(a \to 0)$ and the big rip $(a \to \infty)$, the gravitational coupling goes to infinity, $G \to \infty$. Choosing again
\be
\label{Lamatan}
\Lambda \equiv \frac{\pi^2}{t_s^2} \hspace{0.2cm} {\rm and} \hspace{0.2cm} a_0=\frac{1}{\sqrt{\Lambda}},
\ee
the Friedmann equation reads 
\be\label{FE03}
H^2 = \frac{1}{a^2} \left(1+\Lambda a^2\right)^2 = \Lambda^2 a^2 + 2 \Lambda + \frac{1}{a^2} ,
\ee
where the first term on the right-hand side scales as phantom matter \cite{phantom}, which drives a big-rip singularity.

\section{Quantum multiverses and the entanglement}
\label{quantum}

Motivated by an idea of Ref.~\cite{Marosek2016} to construct models of parallel universes (the multiverse) which have the same geometrical evolution, but a different evolution of the physical constants, and anticipating quantum effects at the points of the classical singularities and turning points of the evolution, we now extend our considerations of cyclic universes into a quantum domain allowing for the interaction between those parallelly evolving universes. In particular, the doubleverse model of two parallel universes will become a prototype of a quantum entangled pair of universes which can spontaneously be created at some special points of the minisuperspace. Another approach to the problem of constructing cyclic universes and then multiverse models has been developed in Ref.~\cite{piao}.  

\subsection{Wheeler-DeWitt (second) quantization}

Let us now canonically quantize the models being classically depicted in Sect.~\ref{classical}. Taking into account the classical value of the momentum conjugated to the scale factor,
\be\label{CMO}
p_a = - a \frac{d a}{d t} ,
\ee
the Hamiltonian constraint, which can be written as
\be\label{HC01}
p_a^2 - \omega^2(a) = 0 ,
\ee
can easily be derived from the Friedmann equations (\ref{FE02}) and (\ref{FE03}), with
\be\label{OM01}
\omega^2_\text{sin}(a) \equiv  a^2 - \Lambda a^4 .
\ee
for the sinusoidal pulse and
\be\label{OM02}
\omega^2_\text{tan}(a) \equiv \Lambda^2 a^6 + 2 \Lambda a^4 + a^2.
\ee
for the tangential pulse.

By canonically quantizing the classical momentum, $p_a \rightarrow - i \frac{\partial}{\partial a}$, and with an appropriate choice of factor ordering\footnote{A different choice of factor ordering would introduce a mass term in the equation of the generalized harmonic oscillator (\ref{WDW01}). It would not modify either the procedure or the qualitative meaning of the results.}, we arrive at the Wheeler-DeWitt equation
\be\label{WDW01}
\ddot{\phi} + \omega^2 \phi = 0 ,
\ee
where, $\phi\equiv\phi(a)$, is the wave function of the universe and the dot indicates a derivative with respect to the scale factor, i.e. $\dot{\phi}\equiv \frac{d\phi}{d a}$. In (\ref{WDW01}) $\omega^2(a)$ defined by (\ref{OM01}) or (\ref{OM02})  plays the role of the Wheeler-DeWitt potential which is the base for the studies of different scenarios due to the boundary conditions for the wave function \cite{Atkatz,kieferbook,vilenkintunnel,HH}. The solutions of (\ref{WDW01}) corresponding to the Wentzel-Kramers-Brillouin (WKB) approximation are given by 
\be\label{WKB01}
\phi_\pm \propto \frac{1}{\sqrt{2 \omega}}e^{\pm i S} ,
\ee
where, $\dot{S} = \omega$. For the sinusoidal pulse, we then get
\be
S = \int da \, \omega_\text{sin}(a) = -\frac{\left( 1 - \Lambda a^2 \right)^\frac{3}{2}}{3 \Lambda}  .
\ee
Let us notice that for $a\in(0,a_0)$, with $a_0\equiv \frac{1}{\sqrt{\Lambda}}$, the WKB wave function (\ref{WKB01}) would represent a Lorentzian (classical) universe, whereas for the value $a>a_0$, the wave function represents the exponential decay of the Euclidean regime or the quantum barrier, as it was expected.

\begin{figure}
\centering
\includegraphics[width=0.5\textwidth]{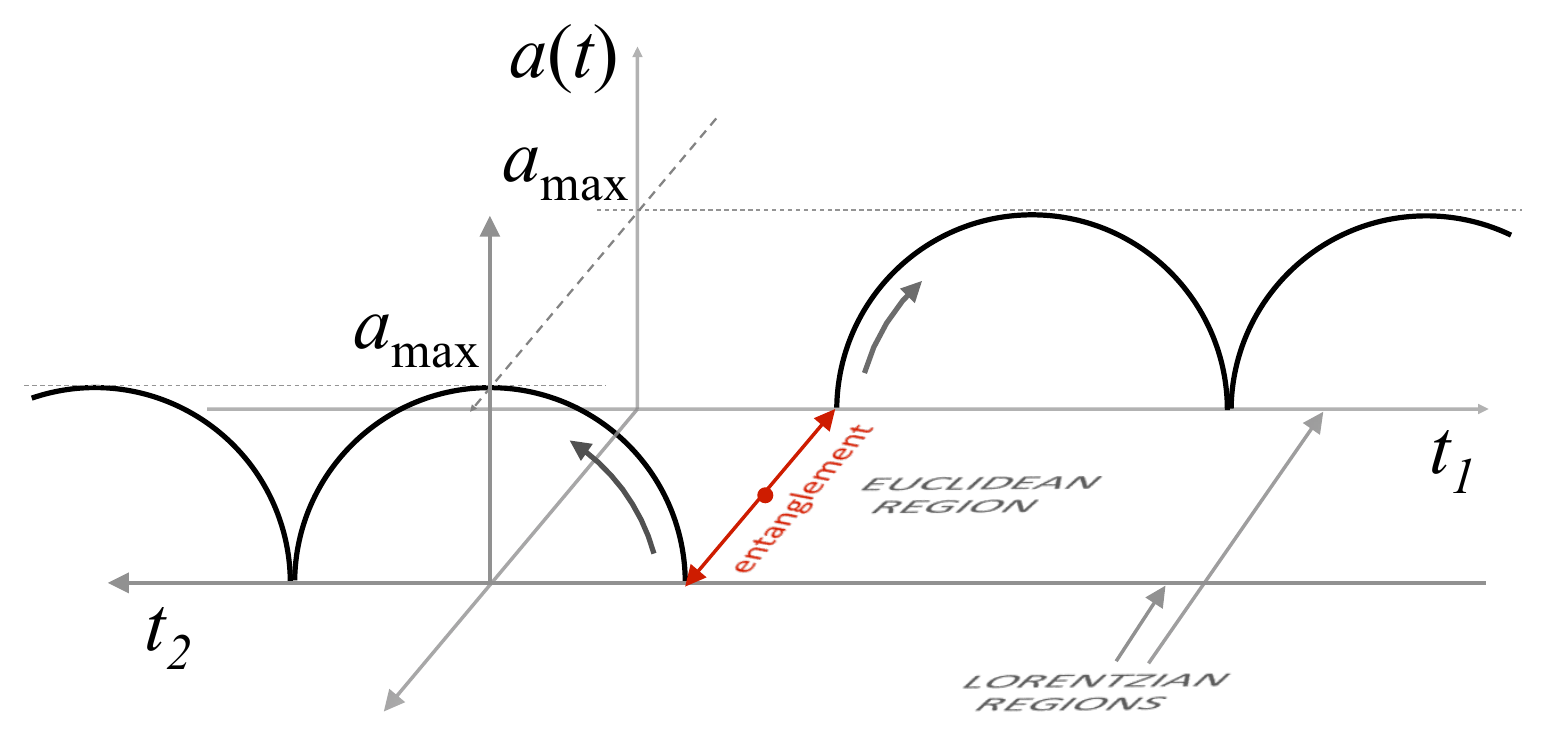}
\caption{Creation of cyclic universes in entangled pairs (sinusoidal pulse).}
\label{fig3}
\end{figure}

The two signs in the exponent of (\ref{WKB01})  correspond to two different branches of the universe being considered. Let us notice that the eigenvalue of the momentum for the WKB solutions (\ref{WKB01}) is given, at first order, by
\be\label{QMO}
\hat{p} \phi_\pm \equiv -i \frac{\partial \phi_\pm}{\partial a} \approx \pm \dot{S} \phi_\pm = \pm \omega \phi_\pm ,
\ee
and in the semiclassical limit it must be highly peaked around the classical value $p_a$, given by Eq. (\ref{CMO}). Then, $a \frac{d a}{d t} \approx \mp \omega(a)$, for the two signs given in Eq. (\ref{WKB01}), and thus
\be
\frac{da}{dt} = \pm \sqrt{1-h^2 a^2} ,
\ee
where $h^2 \equiv \Lambda$, and $\Lambda$ is given in Eq. (\ref{Lama0}). We thus obtain two classical branches, one with a scale factor given by
\be\label{branch01}
a(t) = \frac{1}{h} \sin [h(t-t_0)] ,
\ee
and the other with scale factor given by
\be\label{branch02}
a(t) = \frac{1}{h} \sin [h (t_0 - t)] .
\ee
They are related by the time symmetry, $t \rightarrow - t$ ($t_0 \rightarrow -t_0$), so they appear to be the same universe for any internal observer, provided that the universes are created in entangled pairs (see Fig.~\ref{fig3}) and that the time variables of the observers follow an antipodal-like symmetry \cite{Linde1988, RP2014}. Before reaching the big crunch singularities, which are avoided by the effects of the varying gravitational constant (\ref{VC01}) (see Ref.~\cite{Marosek2016}), one branch of the universe can undergo a quantum transition to the other branch universe, appearing there as a newborn universe, forming thus a continuous and cyclic multiverse.

\begin{figure}
\centering
\includegraphics[width=0.35\textwidth]{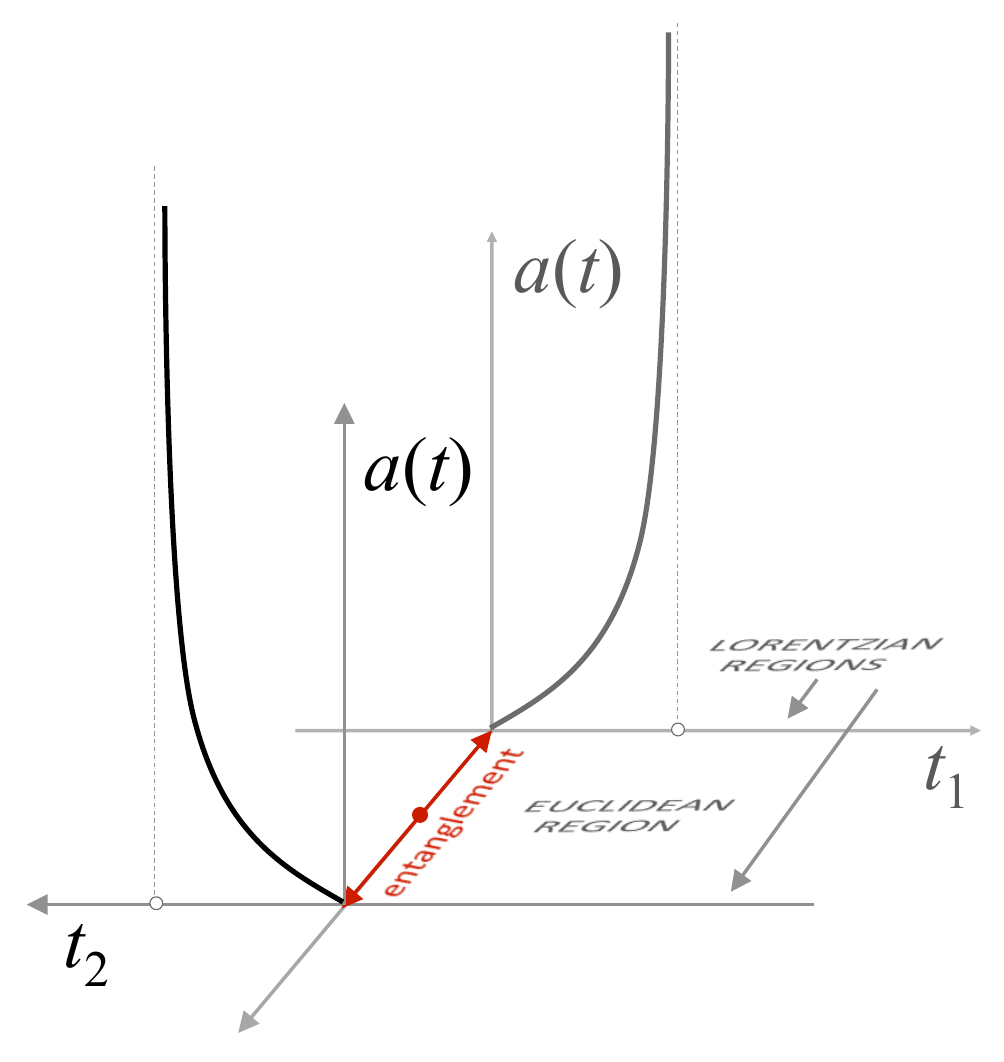}
\caption{Creation of cyclic universes in entangled pairs (tangential pulse).}
\label{fig4}
\end{figure}

For the tangential pulse, we arrive at
\be
S = \int da \, \omega_\text{tan}(a) = \frac{1}{4}\,a^2\left(2 + \Lambda a^2 \right).
\ee
Following a similar reasoning to that made for the sinusoidal pulse, the evolution of the two branches that correspond to the plus and minus signs of $\phi_\pm$ in Eq.~(\ref{WKB01}) is given now by
\be
\frac{d a}{dt} = \pm \left( h^2 a^2 + 1 \right) ,
\ee
with, $h^2 \equiv \Lambda$, where $\Lambda$ is given by Eq.~(\ref{Lamatan}). We thus obtain
\be\label{branch11}
a(t) = \frac{1}{h} \tan [h(t -t_0)] ,
\ee
and
\be\label{branch12}
a(t) = \frac{1}{h} \tan [h (t_0 -t)] ,
\ee
for the two branches of the tangential pulse. They are depicted in Fig.~\ref{fig4}. 

We can now describe the creation of cyclic universes in entangled pairs. The universes are not singular at the value $a=0$ because the varying constants make finite the value of the mass density and pressure at the turning points \cite{Marosek2016}. However, it is expected that quantum effects would become dominant as we approach the value $a=0$. Furthermore, if quantum fluctuations of the wave function of the universe are considered \cite{RP2014}, then, a minimum value $a_\text{min}$ appears, below of which no real solution can be found. In this classically forbidden region, double Euclidean instantons can be created giving rise, in the Lorentzian regime, to an entangled pair of universes whose quantum   states  are quantum-mechanically correlated (see, Figs.~\ref{fig3}--\ref{fig4}). The antipodal symmetry \cite{Linde1988, RP2014} makes an observer living in the universe with time variable $t_1$ to consider her branch as the expanding branch and the preceding one as the contracting 
 branch. However, for the observer of the universe with time variable $t_2$ they are the other way around, actually. Both observers are thus initially living in an expanding universe and the two branches can be combined to form a universe that is classically indistinguishable from the picture depicted in Figs.~\ref{fig1} and \ref{fig2}, respectively.

\subsection{Third quantization}

The creation of universes in entangled pairs can properly be described in the framework of the third quantization, which parallels the formalism of a quantum field theory of the wave function of the universe propagating along the (mini-)superspace. In that framework, creation and annihilation operators can formally be defined much in a similar way to how is done in a usual quantum field theory. Let us first notice that (\ref{WDW01}) can be considered as the wave equation of a scalar field (the wave function of the universe, $\phi$) that can be obtained from the Hamilton equations of the following (third-quantized) Hamiltonian \cite{Strominger,salva2010,salva2014}
\be\label{3H01}
{\rm H} = \frac{1}{2} P_\phi^2 +\frac{\omega^2(a)}{2} \phi^2 ,
\ee
where, $P_\phi \equiv \dot{\phi}$, and $\omega$ is given by (\ref{OM01}) or (\ref{OM02}), for the sinusoidal and the tangential pulse, respectively. In the third quantization formalism the wave function of the universe, $\phi$, and the conjugate momentum, $P_\phi$, are promoted to be operators in a similar way as it is done in a quantum field theory. The wave function operator can be written, in the Heisenberg picture, as
\be\label{WF03}
\hat{\phi}(a) = \frac{1}{\sqrt{2 \omega}} e^{i S(a) } \hat{b}_+ +  \frac{1}{\sqrt{2 \omega}} e^{- i S(a) } \hat{b}_-^\dag ,
\ee 
where, $\hat{b}_+ \equiv \hat{b}_+(a_\text{min})$ and $\hat{b}_-^\dag\equiv \hat{b}_-^\dag(a_\text{min})$, are constant operators given at some initial value, $a=a_\text{min}$, at which the universes are created. For the sinusoidal pulse, $\hat{b}_-$ and $\hat{b}_-^\dag$ would represent the annihilation and creation operators, respectively, of the branches of the universe given by (\ref{branch01}), and $\hat{b}_+$ and $\hat{b}_+^\dag$ are the annihilation and creation operators, respectively, of the branches of the universe given by (\ref{branch02}), both evaluated at the constant value, $a=a_\text{min}$.  Analogously for the tangential pulse, $\hat{b}_-$ and $\hat{b}_-^\dag$ would represent the annihilation and creation operators, respectively, of the branches of the universe given by (\ref{branch11}), and $\hat{b}_+$ and $\hat{b}_+^\dag$ are the annihilation and creation operators, respectively, of the branches of the universe given by (\ref{branch12}), both evaluated at the constant value, $a=a_\text{min}$. The branches are  created in entangled pairs because of the quantum symmetry of the Wheeler-DeWitt equation (\ref{WDW01}) with respect to the value $\pm\omega$ of the classical branches,  quantum-mechanically represented by $\phi_\pm$. This is formally similar to the creation of particles in entangled pairs with opposite directions in a quantum field theory because the symmetry of the wave equation with respect to the values $\pm k$ of the momentum of the particles.

The vacuum state of the $(b_\pm, b^\dag_\pm)$ representation is given by the state, $|0_+,0_-\rangle$. However, it is not a stable vacuum  because of the scale-factor dependence of the frequency $\omega(a)$. Similarly to what is done in a quantum field theory of a scalar field that propagates in a curved spacetime, where it is imposed that the vacuum state should be stable (i.e.~with no particle creation) along a geodesic, we can impose here the boundary condition for the proper  representation for the vacuum state of the minisuperspace that it has to be stable under the evolution of the universe along a geodesic of the minisuperspace. The minisuperspace that we are considering here is the most simplified one and it is just formed by the scale factor as the configuration variable. However, in more detailed cosmological models, the minisuperspace is formed by the scale factor and the scalar field, $\varphi$, that represents the energy-matter content of the universe. Then, a geodesic of the minisuperspace is precisely the path given by the classical relation, $\varphi = \varphi(a)$. The boundary condition that the cosmological vacuum is stable along the geodesic of the minisuperspace means that it is stable under the classical evolution of the universes, i.e., once the multiverse is in the state\footnote{Or more exactly in a superposition state $\sum c_N |N\rangle$.} $|N\rangle$ of the invariant representation for some value $a_0 > a_\text{min}$, then, it will remain in that state at any other value of the scale factor $a(t)$ along the evolution of any universe.

The proper representation for the vacuum state of the multiverse is then given by an invariant representation. For the generalized harmonic oscillator (\ref{WDW01}), it can be given by\footnote{This invariant representation is not unique, see for instance Ref. \cite{Kim2001}. Moreover, the operators $c$ and $c^\dag$ are given in the Schr\"{o}dinger representation, i.e., $\phi = \frac{1}{\sqrt{2\omega}} (b_+ + b_-^\dag)$ and $P_\phi = i \sqrt{\frac{\omega}{2}} (b_-^\dag - b_+)$.} \cite{Lewis1969, salva2010, salva2014}
\begin{eqnarray}\label{IR01}
c_+ &=& \sqrt{\frac{1}{2}} \left( \frac{1}{R} \phi + i(R P_\phi - \dot{R} \phi ) \right) , \\ \label{IR02}
c^\dag_- &=& \sqrt{\frac{1}{2}} \left( \frac{1}{R} \phi - i(R P_\phi - \dot{R} \phi ) \right) ,
\end{eqnarray}
where $R = \sqrt{\phi_1^2 +\phi_2^2}$, with $\phi_1$ and $\phi_2$ being two real solutions of (\ref{WDW01}) satisfying\footnote{More generally, $R$ can be given by $R=\sqrt{A \phi_1^2 + B \phi_2^2 + 2 C \phi_1 \phi_2}$, where $AB -C^2 = W^{-2}$, being $W$ the Wronskian of the two particular solutions $\phi_1$ and $\phi_2$, i.e. $W = \phi_1 \dot{\phi}_2 - \dot{\phi}_1 \phi_2$ (see Ref. \cite{Leach1983}).} 
\be
\phi_1 \dot{\phi}_2 - \dot{\phi}_1 \phi_2 = 1 .
\ee
However, in terms of the invariant representation (\ref{IR01})--(\ref{IR02}), the Hamiltonian (\ref{3H01}) reads
\be\label{H21}
H = H_0^- + H_0^+ + H_I ,
\ee
where
\be
H_0^\pm = \Omega(a) \left( c_\pm^\dag c_\pm + \frac{1}{2}\right)  , 
\ee
and,
\be
H_I =  \gamma(a) c_+^\dag c_-^\dag + \gamma^* c_+ c_- ,
\ee
with
\begin{eqnarray}\label{OM01}
\Omega(a) &=& \frac{1}{4} \left( \frac{1}{R^2} + R^2 \omega^2 + \dot{R}^2 \right) , \\
\gamma(a) &=& -\frac{1}{4} \left\{ \left( \dot{R} + \frac{i}{R} \right)^2 + \omega^2 R^2 \right\} .
\end{eqnarray}
The Hamiltonian (\ref{H21}) can be interpreted as the Hamiltonian of two interacting universes with a Hamiltonian of interaction given by $H_I$. The picture is then the following. A hypothetical external observer moving along a geodesic of the minisuperspace would perceive it in the vacuum state. The only universes that would be created, from this point of view, would be virtual universes created in entangled pairs due to the symmetry of the quantum components of classical solutions given by, $p_a = \pm\omega$. The entanglement between the universes of each entangled pair can be seen as a non-local interaction given by $H_I$ that goes to zero as the entanglement disappears. In that limit, the invariant representation becomes the diagonal representation of the Hamiltonian (\ref{3H01}), 
\begin{eqnarray}\label{DR01}
b_+(a) &=& \sqrt{\frac{\omega}{2 }} \left(\phi + \frac{i}{\omega} P_\phi\right) , \\  \label{DR02}
b_-^\dag(a) &=& \sqrt{\frac{\omega}{2 }} \left(\phi - \frac{i}{\omega} P_\phi\right) ,
\end{eqnarray}
with $\omega \equiv \omega(a)$ given by (\ref{OM01}) or (\ref{OM02}) for the sinusoidal and the tangential pulse, respectively. For the value $a=a_\text{min}$, it is the Schr\"{o}dinger picture of the representation (\ref{WF03}). However, the representation (\ref{DR01})--(\ref{DR02}) can  represent the state of the universe for any other value of the scale factor. For instance, it may represent the quantum state of an evolved universe like ours, with $a \gg a_\text{min}$, with inhabitants living on a planet there. For such an observer, i.e.~for an internal observer, $b(a)$ and $b^\dag(a)$ would not describe  annihilation and creation of universes because these observers can only perceive their own universe. Instead, they would  represent the annihilation and creation of quantum modes of the general quantum state of their single universes.

\section{Thermodynamics of entanglement}
\label{thermodynamics}

\subsection{General framework}

The scenario is then the following: the multiverse is in the vacuum state, which is quantum-mechanically described by the ground state of the invariant representation of the minisuperspace, $| 0_+ 0_-\rangle_c$. Given that the ground state  $| 0_+ 0_-\rangle_c$ is a pure state, its entropy is zero, and because it follows a unitary evolution in the minisuperspace, the entropy is constantly zero. From this point of view, therefore, there would be no arrow of time in the multiverse as it corresponds to a steady system. However, it is reasonable to think that the real evolution and the appearance of a physical arrow of time would only make sense in the context of a single universe for an internal observer. Such an arrow of time could be given by the entropy of entanglement of each single universe, which not only is not zero but it evolves with respect to the value of the scale factor, and provides a relationship between the physical and the mathematical arrows of time in each individual 
 universe, as it corresponds to the point of view of an internal observer who does not see the rest of the multiverse.

Let us therefore consider the ground state of the invariant representation, $| 0_+ 0_-\rangle_c$.  In terms of the diagonal representation, $(\hat{b}_+,\hat{b}_-)$, which  would represent the state of the universe for an internal observer, it is given by\footnote{The formalism parallels that given in Ref.~\cite{salva2014}.}
\be\label{VS01}
| 0_+ 0_-\rangle_c = \frac{1}{|\alpha|} \sum_{n=0}^\infty \left( \frac{|\beta| }{ |\alpha|} \right)^n |n_-, n_+\rangle_b ,
\ee
where $|n_-, n_+\rangle_b $ are the entangled mode states of the diagonal representation given by (\ref{DR01})--(\ref{DR02}), and $\alpha$ and $\beta$ are the Bogoliubov coefficients that relate both representations, i.e.
\begin{eqnarray}\label{BO01}
\hat{c}_- &=& \alpha \hat{b}_- - \beta \hat{b}_+^\dag , \\ \label{BO02}
\hat{c}_-^\dag &=& \alpha^* \hat{b}_-^\dag - \beta^* \hat{b}_+ ,
\end{eqnarray}
with, $|\alpha|^2 - |\beta|^2 = 1$. The plus and minus signs correspond to the two branches of the universe. We can now obtain the quantum state of a single universe of the entangled pair in the $(\hat{b}_+,\hat{b}_-)$ representation by tracing out the degrees of freedom of the partner universe. In the formalism of the density matrix
\be\label{TO}
\rho_{-} = {\rm Tr}_+ \rho \equiv \sum_{n=0}^\infty {}_b\langle n_+ | \rho | n_+\rangle_b,
\ee
where
\begin{eqnarray}\nonumber
\rho &=& | 0_+ 0_-\rangle_c \langle 0_+ 0_-| \\  \label{r1}
&=& \frac{1}{|\alpha|^2} \sum_{n,m} \left( \frac{|\beta|}{|\alpha|} \right)^{n+m} |n_-,n_+\rangle_b\langle m_-,m_+| ,
\end{eqnarray}
where (\ref{VS01}) has been used. The result of the trace operation in (\ref{TO}) is typically a thermal state, given by \cite{salva2014}
\begin{eqnarray}\nonumber
\rho_{-} &=& \frac{1}{|\alpha|^2} \sum_{n,m,l}  \left( \frac{|\beta|}{|\alpha|} \right)^{n+m}  \langle l_+ | m_+\rangle | n_- \rangle_b\langle n_-| \langle m_+ | l_+\rangle \\ \nonumber
&=& \frac{1}{|\alpha|^2} \sum_{n}  \left( \frac{|\beta|}{|\alpha|} \right)^{2n}  |n_-\rangle_b\langle n_-| \\ \nonumber
&=& \frac{1}{|\alpha| |\beta|} \sum_{n}  \left( \frac{|\beta|}{|\alpha|} \right)^{2n + 1}  |n_-\rangle_b\langle n_-| \\ \label{RS02}
&=& \frac{1}{Z} \sum_n e^{-\frac{\omega}{T} (n +\frac{1}{2})} |n_-\rangle_b\langle n_-| ,
\end{eqnarray}
where, $Z^{-1} = 2 \sinh\frac{\omega}{2T}$, with
\be
\label{T01}
T\equiv T(a) = \frac{\omega(a)}{2 \ln\coth r} ,
\ee
where
\be
\tanh r \equiv \frac{|\beta|}{|\alpha|} ,
\ee
with $r$ playing the role of the entanglement parameter \cite{nakagawa16,Baskal16}.
Moreover, in order to obtain (\ref{RS02}), we have used
\begin{eqnarray} \nonumber
Z^{-1} &=& 2 \sinh\frac{\omega}{2T} = 2 \sinh\ln\coth r \\
&=& \coth r - \tanh r = \frac{1}{\sinh r \cosh r} .
\end{eqnarray}
In fact, we have derived the corresponding thermal state that represents the state of a single universe of the entangled pair for an internal observer from the zero entropy vacuum state of the superspace of an external observer. The quantum entropy or \emph{entropy of entanglement} of the universe can now be easily obtained from (\ref{RS02}) \cite{horodeckiRMP,nishioka,Casini,nakagawa16,Baskal16,Sengupta16}. It is given by the von Neumann entropy
\be
S(\rho) = - \rm{Tr}\left( \rho \ln \rho \right) ,
\ee
applied to the thermal state $\rho_-$, and yields \cite{salva2014}
\begin{equation}
\label{eq68}
S_\text{ent}(a) = \cosh^2 r \, \ln \cosh^2 r - \sinh^2 r \, \ln \sinh^2 r .
\end{equation}
The dependence of the entropy of entanglement on the scale factor means that the evolution of each single universe is no longer unitary due to the non-local interaction that produces the entanglement. The evolution of an entangled pair, however, is unitary and so there is no information paradox for an external observer.

It is also worth noticing that the same value of entropy would be obtained for the partner universe, i.e. $S_\text{ent}(\rho_+) = S_\text{ent}(\rho_-)$, satisfying the subadditivity of entropy theorem \cite{Araki1970}
\be
S(\rho) \leq S(\rho_-) + S(\rho_+) = 2 S(\rho_\pm),
\ee
where the inequality is saturated whenever $\rho_+$ and $\rho_-$ correspond to two uncorrelated (classical) universes with
\be
\frac{d S_+}{d a } = \frac{d S_-}{da} ,
\ee
and $S_\pm \equiv S(\rho_\pm)$. A change of the entropies with respect to the internal time variables is
\be
\frac{d S_+}{ d t_1} = \frac{d S_-}{dt_2} \Rightarrow \frac{d S_+}{ d t_{1,2}} = - \frac{d S_-}{dt_{1,2}}  ,
\ee
provided that the time variables $t_1$ and $t_2$ of the branches are related by the antipodal symmetry commented earlier after Eq. (\ref{branch02}).

Other parameters of quantum thermodynamics can be defined as well \cite{salva2014} (see also, Refs. \cite{Alicki2004, Gemmer2009}). The mean value of the Hamiltonian 
\begin{equation}
\hat{H}_- = \omega \left(\hat{b}_-^\dag \hat{b}_- + \frac{1}{2} \right), 
\end{equation}
turns out to be
\begin{equation}\label{E31}
E_-(a) \equiv \langle \hat{H}_- \rangle =  {\rm Tr} \hat{\rho}_- \hat{H}_- =  \omega \left(\langle \hat{N}(a) \rangle + \frac{1}{2}\right) ,
\end{equation}
with
\be\label{N31}
\langle \hat{N}(a) \rangle = \sinh^2 r .
\ee
Changes in the quantum informational analogues of heat and work are \cite{salva2014}
\begin{eqnarray}
\delta W_- &=& {\rm Tr} \left(\hat{\rho}_- \frac{d \hat{H}_-}{d a} \right) = \frac{\partial \omega}{\partial a} \left(\langle \hat{N}(a) \rangle + \frac{1}{2}\right) , \\ \label{eq6332}
\delta Q_- &=& {\rm Tr} \left(\frac{d \hat{\rho}_-}{d a} \hat{H}_- \right) = \omega \frac{\partial \langle \hat{N}(a) \rangle}{\partial a} .
\end{eqnarray}
It can easily be checked that the first law of thermodynamics is satisfied, i.e. $dE_- = \delta W_- + \delta Q_-$. It can also be checked that the production of entropy is zero, 
\begin{equation}\label{eq6333}
\sigma = \frac{d S_\text{ent}}{da} - \frac{1}{T} \frac{\delta Q}{d a} = 0 ,
\end{equation}
with $T$ being defined in Eq. (\ref{T01}). It thus corresponds to a reversible process. This was expected because no dissipative process has been taken into account. It means that the entanglement alone does not provide us with an arrow of time because the evolution leading to an increasing value of the scale factor or that leading to a decreasing value are both allowed. However, if local dissipative processes are taken into account, then, the production of entropy must necessarily be positive, i.e. $\sigma \geq 0$, making the evolution of the universe irreversible. Let us notice that by local processes in the context of the multiverse we mean any process that may happen inside a single universe like, for instance, the creation of cosmic structures or even customary \emph{non-local} processes in the context of the spacetime non-locality of quantum mechanics, i.e.~any process that is not correlated with any other process of the partner universe.

\subsection{The sinusoidal pulse -- entanglement quantities}

\begin{figure}
\includegraphics[width=0.5\textwidth]{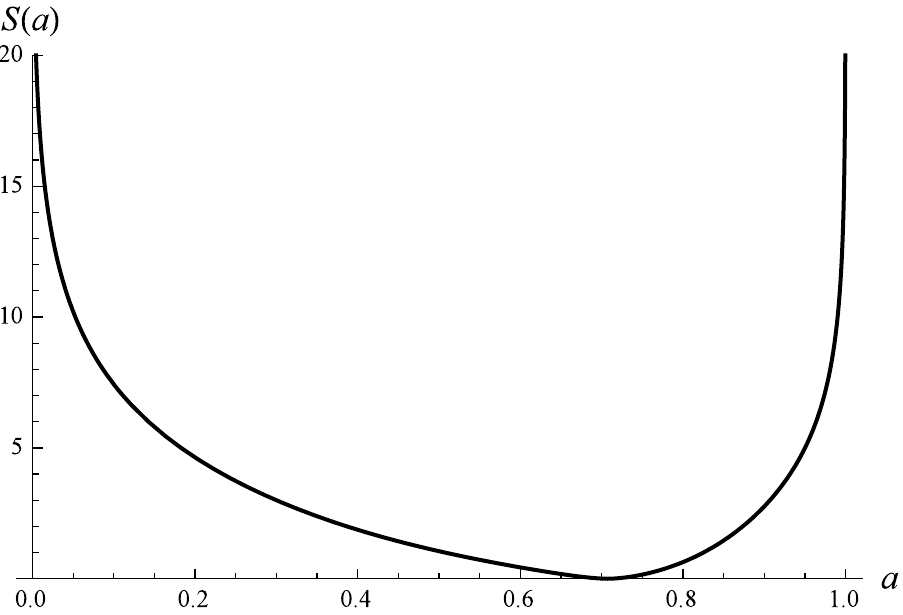}
\caption{The entropy of entanglement for the sinusoidal pulse plotted in terms of the scale factor, where $a=1$ corresponds to $a_\text{max}$.}\label{Ssina-plot}
\end{figure}

\begin{figure}
\includegraphics[width=0.5\textwidth]{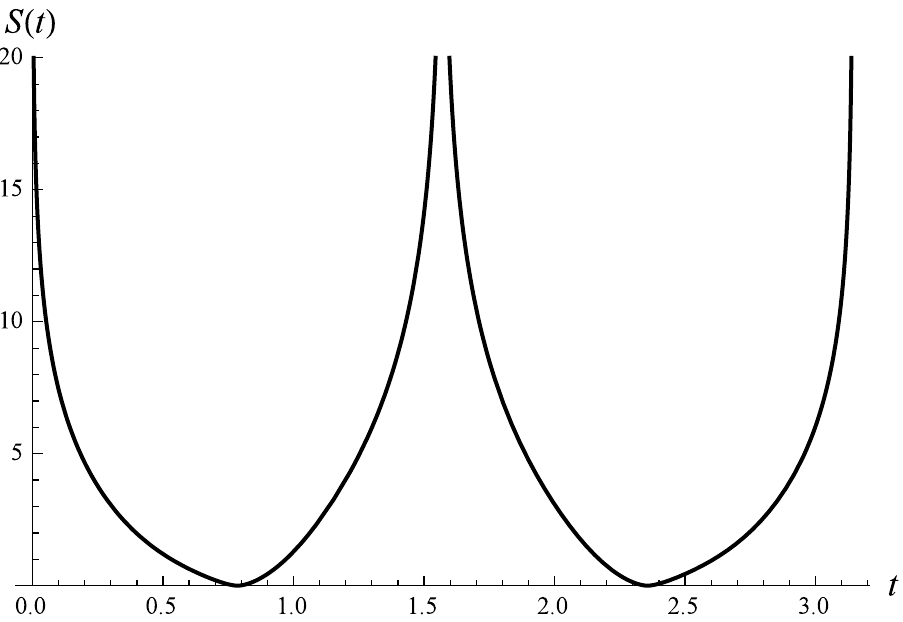}
\caption{The entropy of entanglement for the sinusoidal pulse plotted in terms of the cosmic time, where $t=\pi/2$ corresponds to the point of maximum expansion and $t=\pi$ to the point of the big crunch.}\label{Ssint-plot}
\end{figure}

We can now compute the entropy of entanglement for the cyclic multiverse considered in Sections \ref{classical} and \ref{quantum}. First, one derives $\phi$ and $P_\phi$ from (\ref{DR01}) and (\ref{DR02}), then inserts them into (\ref{IR01}) and (\ref{IR02}), in order to get that the values of $\alpha$ and $\beta$ in (\ref{BO01}) and (\ref{BO02}) are given by
\begin{eqnarray}
\alpha &=& \frac{1}{2} \left( \frac{1}{R\sqrt{\omega}} + R \sqrt{\omega} - \frac{i \dot{R}}{\sqrt{\omega}} \right) , \\
\beta &=& - \frac{1}{2} \left( \frac{1}{R\sqrt{\omega}} - R \sqrt{\omega} - \frac{i \dot{R}}{\sqrt{\omega}} \right) ,
\end{eqnarray}
with $R=\sqrt{\phi_1^2 + \phi_2^2}$, being $\phi_1$ and $\phi_2$ two real solutions of the Wheeler-DeWitt equation (\ref{WDW01}). Considering linear combinations of the WKB solutions (\ref{WKB01}), a natural choice for $\phi_1$ and $\phi_2$ is
\begin{eqnarray}
\phi_1 &=& \frac{1}{\sqrt{\omega}} \cos S , \\ 
\phi_2 &=&  \frac{1}{\sqrt{\omega}} \sin S ,
\end{eqnarray}
which yields $R\sqrt{\omega} = 1$, and
\begin{eqnarray}\label{A01}
\alpha &=& 1+  \frac{i \dot{\omega}}{4\omega^2} , \\ \label{B01}
\beta &=& - \frac{i \dot{\omega}}{4\omega^2}  ,
\end{eqnarray}
with $|\alpha|^2-|\beta|^2=1$, and where $\dot{R} = -\frac{1}{2}\dot{\omega}\omega^{-\frac{3}{2}}$ has been used. Then,
\begin{equation}
\tanh r = \frac{|\beta|}{ |\alpha|} = \frac{\dot{\omega}}{\sqrt{16\omega^4 + \dot{\omega}^2}} = 
\frac{1}{\sqrt{1 + \left(\frac{4 \omega^2}{\dot{\omega}}\right)^2}} ,
\end{equation}
with, $\dot{\omega}\equiv \frac{d\omega}{d a}$, and $\omega(a)$ given by (\ref{OM01}), so that 
\begin{equation}
\label{qdef}
\tanh r = \frac{1}{\sqrt{1 + 16a^4\frac{\left(1 - \Lambda a^2 \right)^3}{\left(1 - 2 \Lambda a^2 \right)^2}}} \equiv q.
\end{equation}
Note that $q=1$ at zeros of the Wheeler-DeWitt potential (\ref{OM01}) present at $a=0$ and 
$a_\text{max} = 1/\sqrt{\Lambda}$, while $q=0$ at its maximum for $a_\text{c} = \sqrt{2\Lambda}$ \cite{Atkatz}. The temperature of entanglement (\ref{T01}) and the entropy of entanglement (\ref{eq68}) are both measures of the rate of entanglement between the universes and can be rewritten using (\ref{qdef}) as 
\beq\label{Tsin}
T &=& - \frac{a \sqrt{1 - \Lambda a^2}} {2 \ln{q}} ,\\ \label{Ssin}
S &=& \frac{1}{1-q^2} \ln{\left[ \frac{1}{1-q^2} \right]} - \frac{q^2}{1-q^2} \ln{\left[ \frac{q^2}{1-q^2} \right]} .
\eeq
The entropy is plotted in Fig.~\ref{Ssina-plot} in terms of the value of the scale factor $a$, and using Eqs.~(\ref{branch01}) and (\ref{branch02}), it is depicted in Fig.~\ref{Ssint-plot} in terms of the cosmic time $t$. It can be checked that the entanglement is maximum -- in fact, it goes to infinity -- for both the smallest value of the scale factor and also for the maximum value of the scale factor -- the turning point of expansion at $a_\text{max} = \frac{1}{\sqrt{\Lambda}}$. 


Entanglement is usually associated to non-locality. However, there is no need for a common spacetime between the universes of the multiverse. Therefore, the question about locality or non-locality has to be extended in the quantum multiverse to the independence or the interdependence, respectively, of the quantum states of the universes. On the other hand, entanglement is also interpreted as a sharp quantum effect having no classical counterpart. This is so in the sense that the probability distribution of the number of particles in an entangled state may violate certain classical inequalities \cite{Reid1986}. However, we have presented here an example of quantum entanglement between otherwise classical universes [let us recall that the momentum (\ref{QMO}) is highly peaked around the classical value (\ref{CMO}), giving rise to the (semi)-classical branches (\ref{branch01})--(\ref{branch02}) and (\ref{branch11})--(\ref{branch12}) for the sinusoidal and the tangential pulses, respectively]. Therefore, the condition between classicality and entanglement must be revised as well in the context of the quantum multiverse.

In the case of the sinusoidal pulse, the universes originate as an entangled pair. Their quantum states become more and more separable as they evolve towards the value $a_\text{c}$ of the scale factor, where the separability of their quantum states is maximum (their entropy of entanglement is minimum). Afterwards, the entanglement between their states starts growing again to reach a maximum value at the turning point, $a_\text{max}$, where the universes become maximally entangled again. One could then state that at the points of maximum entanglement the quantum effects in the multiverses are expected to be dominant. This is the case, but not because of the maximum amount of entanglement between the universes (we shall see a counterexample in the tangential pulse). The quantum effects become dominant because the proximity of the points $a=0$ and $a=a_\text{max}$ of the configuration space to the classically forbidden region of $a < 0$ and $a > a_\text{max}$, respectively. This is something which fully confirms earlier studies of Refs.~\cite{packets,timearrow,tunnel95,Mithani12,Mithani14}.

\begin{figure}
\centering
\includegraphics[width=0.5\textwidth]{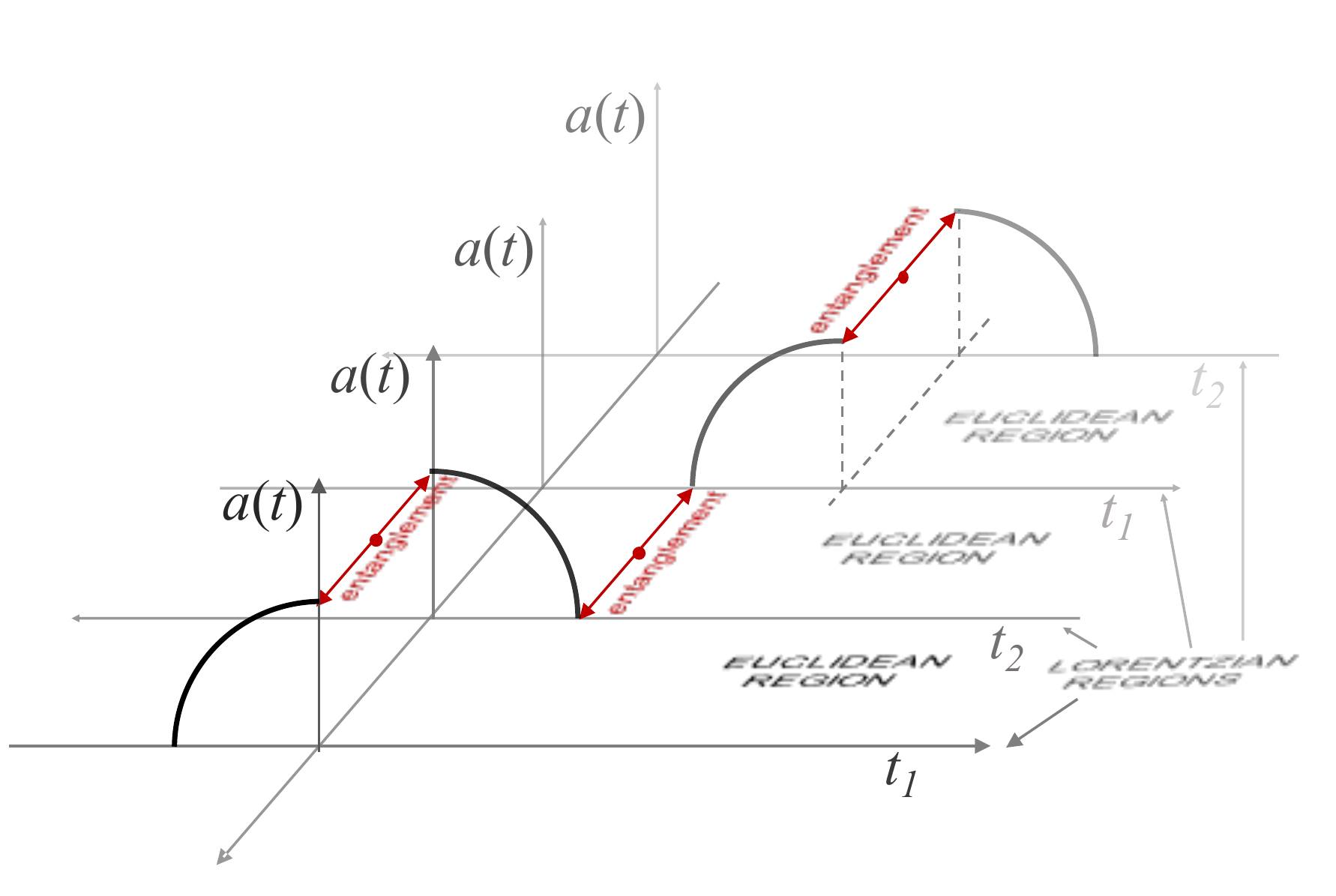}
\caption{Creation of entangled branches of cyclic universes. At the big bang as well as at the maximum expansion the branches they become maximally entangled.}
\label{fig5}
\end{figure}

\subsection{The tangential pulse -- entanglement quantities}

\begin{figure}
\includegraphics[width=0.5\textwidth]{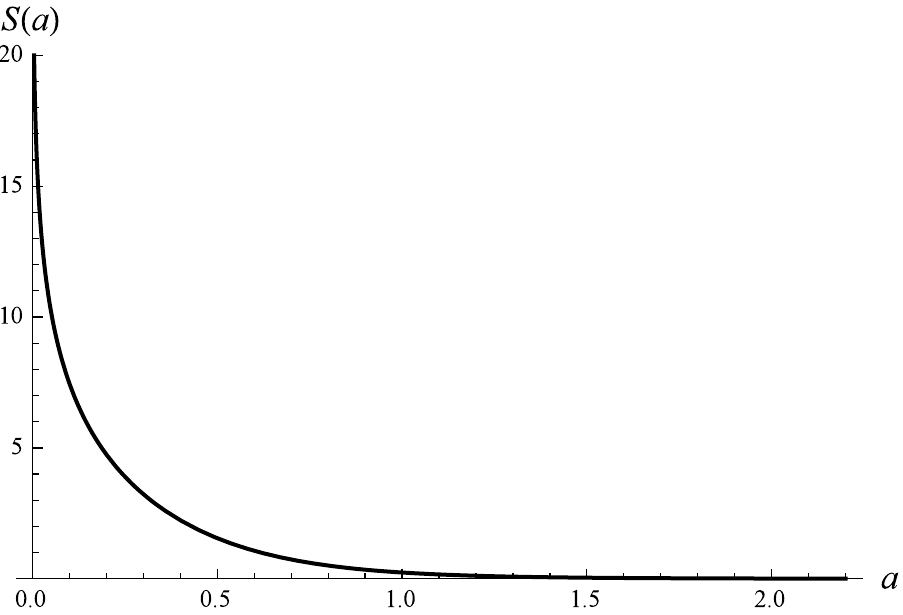}
\caption{The entropy of entanglement for the tangential pulse plotted in terms of the scale factor.}\label{Stana-plot}
\end{figure}

\begin{figure}
\includegraphics[width=0.5\textwidth]{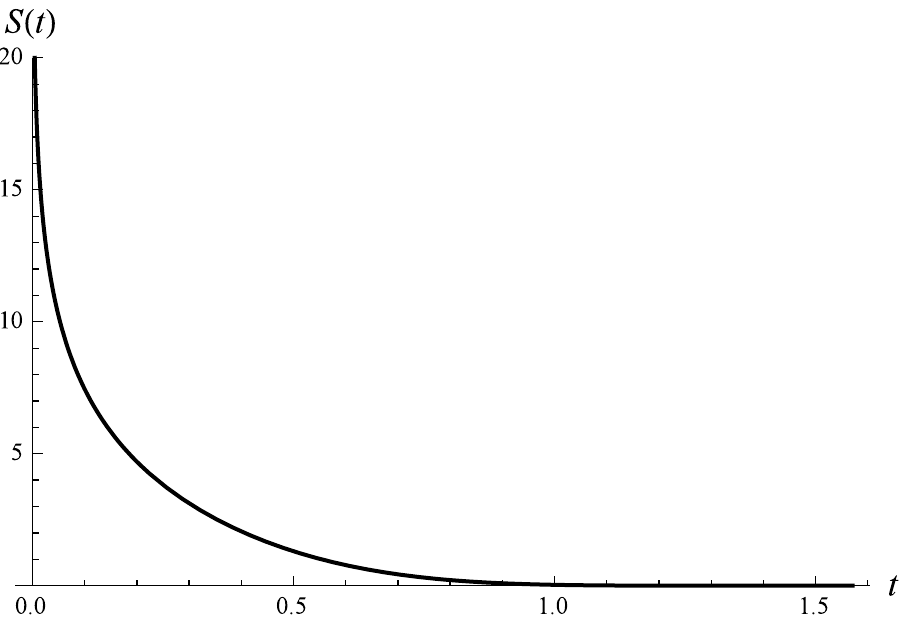}
\caption{The entropy of entanglement for the tangential pulse plotted in terms of the cosmic time, where $t=\pi/2$ corresponds to the point of the big rip.}\label{Stant-plot}
\end{figure}

The same development of the sinusoidal pulse can be made now for the tangential pulse by using the frequency (\ref{OM02}) instead of (\ref{OM01}). In that case, the parameter $q$ turns out to be
\be\label{qtan}
q \equiv \tanh r = \frac{1}{\sqrt{1 + \frac{16 (\Lambda a^3 + a)^4}{(3 \Lambda a^2 + 1)^2}}} .
\ee
Then, the temperature (\ref{Tsin}) now reads
\be
T = - \frac{a (\Lambda a^2 + 1)}{2 \ln q} ,
\ee
and the entropy of entanglement is given by Eq. (\ref{Ssin}) with the value of $q$ given by (\ref{qtan}). The respective plots for $S$ are depicted in Figs.~\ref{Stana-plot} and \ref{Stant-plot}.


It can easily be seen that the entropy of entanglement is maximum -- in fact, again infinite -- at the big bang, but then monotonically decreases and reaches zero at the big rip singularity. This is an example that shows that the amount of entanglement is not correlated, at least in the case of the quantum multiverse, with the classicality of the universes because quantum effects become dominant as the universes approach the big rip singularity \cite{PRD2006}. However, we have shown that the amount of entanglement decreases towards zero as the universes approach the big rip. Their quantum representations become more and more separable and the non-local interaction given by $H_I$ in (\ref{H21}) goes to zero. They can be considered then as individual, non-interacting universes. However, this has nothing to do with the quantum effects of the matter fields that propagate therein. In fact, as it happens in the sinusoidal pulse, these may become dominant because the proximity of the scale factor to a classical forbidden region, which in the case of the tangential pulse is given by $a \rightarrow \infty$ at the value, $t = t_0 + \frac{(2n+1)\pi}{2 h}$ (see Fig. \ref{fig4}).

\begin{figure}
\centering
\includegraphics[width=0.5\textwidth]{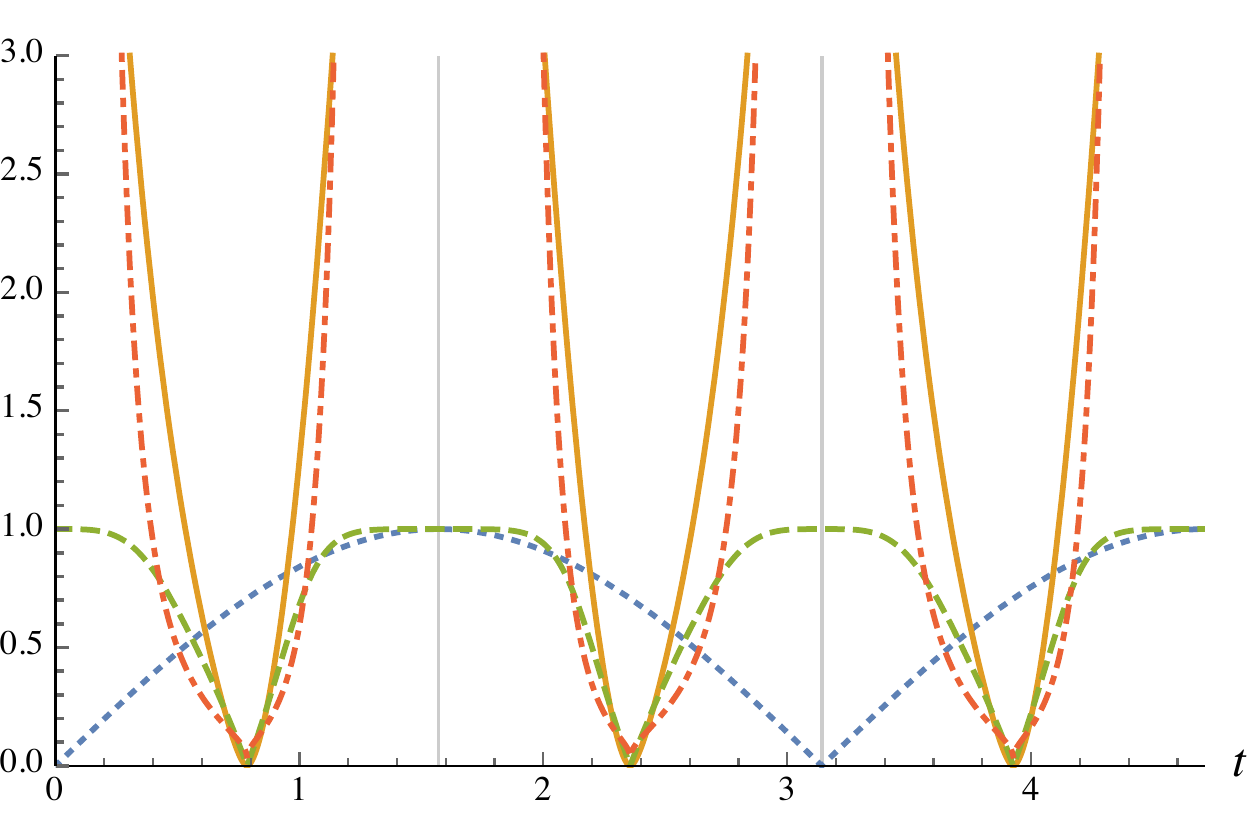}
\caption{Scale factor (blue, dotted), parameter $q$ (green, dashed), entropy of entanglement (yellow, solid line), and temperature of entanglement (red, dot-dashed) for the sinusoidal pulse. Unlike the entropy of entanglement, the parameter $q$ turns out to be a non-divergent measure of the entanglement.}
\label{fig6sin}
\end{figure}

\begin{figure}
\centering
\includegraphics[width=0.5\textwidth]{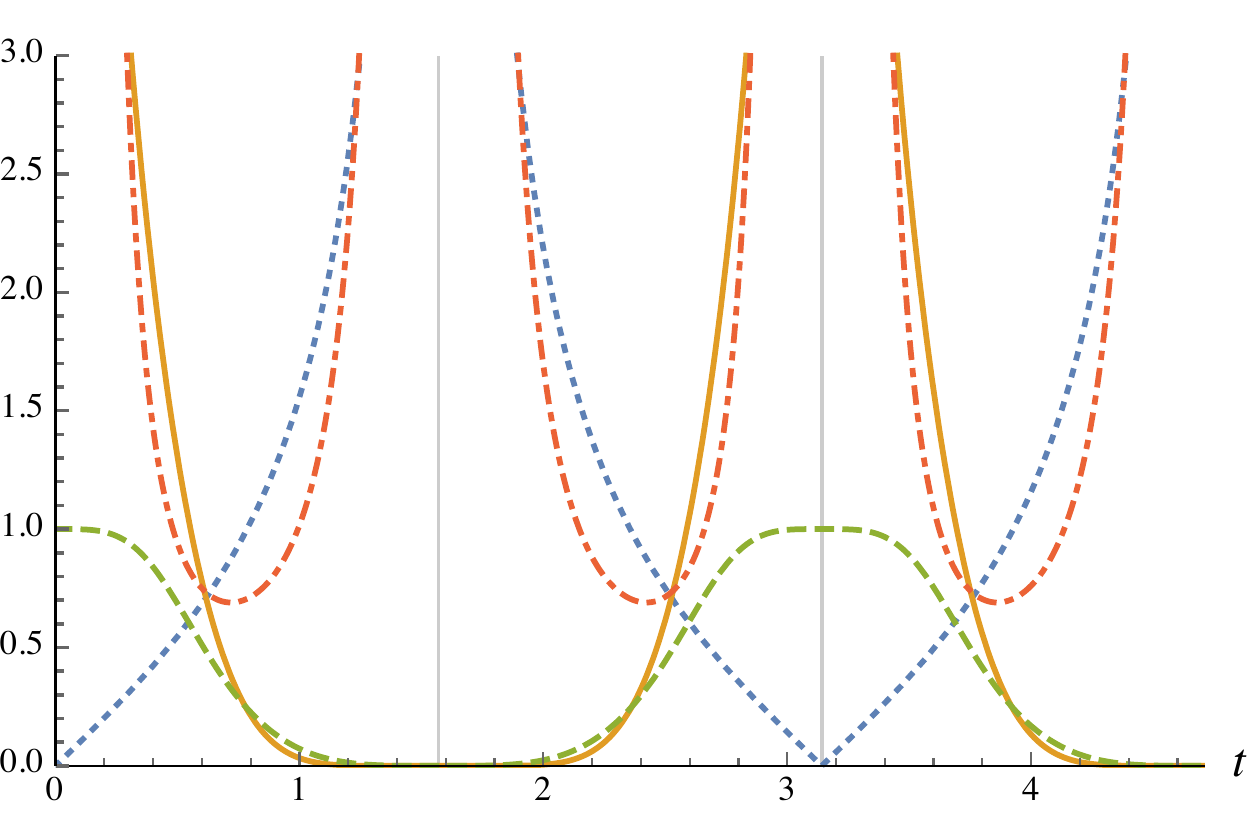}
\caption{Scale factor (blue, dotted), parameter $q$ (green, dashed), entropy of entanglement (yellow, solid line), and temperature of entanglement (red, dot-dashed) for the tangential pulse. The temperature of entanglement might be an indicator of the quantumness of the universes.}
\label{fig6tan}
\end{figure}


\section{Remarks on entanglement thermodynamics and its observational consequences}
\label{observations}

The  third quantization procedure parallels that of a quantum field theory in a curved spacetime. However, there are relevant differences that have to be noticed in order to  understand the analysis of the inter-universal entanglement considered in this and future works.

In a quantum field theory as well as in quantum optics it is customarily assumed that the field at hand (the scalar field or the electromagnetic field, respectively) can be described in terms of a set of quantum oscillators, which under quite general circumstances can represent what we call particles. However, the representation of the fields in terms of particles as individual and independent quantities is not always appropriate or possible. The violation of classical inequalities in quantum optics \cite{Reid1986} precisely shows that this representation fails in some extremal but testable situations. On the other hand, it may seem obvious but is worthy to recall that the concept of a particle is closely attached to the idea of a particle detector. This is a device, external to the field to be measured, that measures pulses of energy because it is weakly coupled to the field. These quantized pulses are what we call particles.

Regarding the quantum description of the spacetime, an example where the interpretation of the wave function of the spacetime in terms of particles is appropriate is the quantum description of the fluctuations of the spacetime in terms of what is called \emph{baby universes}\footnote{In fact, the procedure of third quantization was initially developed to describe this kind of fluctuations (see, Ref.~\cite{Strominger} and references therein).}. As a first approximation, the quantum fluctuations of the spacetime can be described in terms of particle-like pieces of the spacetime of Planck size that branch off from the parent spacetime and propagate therein. The coupling of these baby universes with the matter fields that propagate in the parent spacetime can be described by using a mixed formalism that combines a third quantization formalism for the baby universes and a customary Lagrangian description for the matter fields \cite{Strominger}. Thus, one could use the matter fields as a detector for these fluctuations of the spacetime, and the detected pulses would correspond to baby universes\footnote{Indeed, it should be possible to detect these fluctuations by measuring the coherence properties of matter fields (see Refs.~\cite{PFGD1992a, PFGD1992b, Giddings1998}).}.

Analogously, in the multiverse, for a hypothetical super-observer that would live in the superspace the description of the wave function of the universe in terms of oscillators would not be very different from the one given above. If such an observer would have a \emph{universe-counter}, then, he or she could detect pulses of the wave function of the spacetime that could be interpreted as  universes, being the momentum of these pulses related to the Friedmann equation of the universes, i.e.~related to the type of universes that propagate in the superspace. For that super-observer, the boundary condition that the vacuum of the minisuperspace is quantum-mechanically represented by the ground state of an invariant representation is appropriate because it corresponds to the \emph{no-universe} state for the entire evolution along a geodesic of the superspace. However, if the concept of a particle is observer-dependent so is the concept of a universe and, thus, it happens that the no-universe state for a super-observer may correspond to a different universal state for another possible observer.

For us, as internal observers of our universe, the situation is quite different. We are actually the detector of our own universe and our existence is the result that attests that our universe exists\footnote{In the words of J. Hartle, ``we live in the middle of this particular experiment'', in Ref.~\cite{Hartle1992}, p.~4.}. However, we cannot directly see more universes than ours. Therefore, if any representation has to describe the quantum state of the universe from the point of view of an internal observer, then, the number state of such representation cannot be interpreted as the state representing any number of universes. It does not mean that the wave function of the spacetime cannot be described in terms of the quantum oscillators that would arise from the Fourier transformation in momentum space. It only means that these quantum oscillators cannot be interpreted  as \emph{particles}. They should be interpreted instead as quantum modes that sum up to give a particular state of 
 the universe. Once the mode distribution is known, the variables associated to the field like the energy or the momentum can be computed.

In the present paper we have assumed\footnote{This is, however, not the only possible interpretation.} that  the representation given by Eqs.~(\ref{DR01})--(\ref{DR02}) would represent the state of the universe from the point of view of an internal observer. Then, $|\beta|^2$, with $\beta$ given in (\ref{BO01})--(\ref{BO02}), does not represent the number of universes but the number of the corresponding modes. The same discussion is related to the meaning of the thermal state (\ref{RS02}), which has been obtained by tracing out from the composite state of two entangled universes the degrees of freedom of the partner universe. In principle, a thermal state obtained from an entangled vacuum state is indistinguishable from the thermal distribution of a classical mixture \cite{Partovi2008,Jennings2010}. However, the thermal state (\ref{RS02}) is not the thermal state of any matter field but it is the thermal state of the wave function of the universe from the point of view of {\it an internal observer}. As we have said before, it should not be interpreted in terms of universes but in terms of {\it a thermal distribution of the modes} of the spacetime of our universe.

With that thermal distribution of modes one could compute, for instance, the energy or the temperature of the associated thermal state. However,  the relation that may exist between the thermodynamics of entanglement and the classical description of thermodynamics is not yet clear. It is expected that they are related \cite{Vedral2002, Horodecki2008, Brandao2008}. Even more, the thermodynamics of entanglement is expected to be a quantum generalization of the classical thermodynamics \cite{Plenio1998, Horodecki2008}, so they should coincide in some semiclassical limit. However, what is the appropriate limit to describe this coincidence and what is the exact relation between both formulations is something that is not clear yet.

However, it is a valuable program that if completed, would open the door to a new wide variety of testable experiments, especially in the case of the multiverse because it would provide us with {\it observable imprints} of the multiverse in the properties of our own universe. Let us notice that if the thermal state (\ref{RS02}) and the temperature of entanglement (\ref{T01}) are eventually related to the thermodynamical properties of the universe, then, the energy of entanglement between two or more universes should be accounted for in the Friedmann equation and thus it would have observable consequences in the evolution of the universe \cite{Laura09,kinney16,Laura16a,Laura16b}. Let us recall that in terms of the frequency $\omega$  the Friedmann equation can be written as [see Eqs.~(\ref{CMO}) and (\ref{HC01})]
\be\label{FE21}
\frac{da}{dt} = \frac{\omega}{a} ,
\ee
which gave rise to the solutions (\ref{branch01})--(\ref{branch02}) and (\ref{branch11})--(\ref{branch12}) for the sinusoidal and the tangential pulse, respectively. However, if the state of the states of the universe is given by the thermal state (\ref{RS02}), then, the energy of the thermal state would be given by (\ref{E31}) with (\ref{N31}), and the effective value of the frequency of the ground state would then be given by
\be
E_- = \frac{\omega_\text{eff}}{2} = \omega \left( \sinh^2r + \frac{1}{2} \right) .
\ee
Then, it is expected that the Friedmann equation (\ref{FE21}) will be changed by the effective value of the frequency, i.e
\be\label{FEff01}
\frac{da}{dt} = \frac{\omega_\text{eff}}{a} = \frac{\omega}{a} \left( 1 + 2 \sinh^2r \right)
\ee
The second term in (\ref{FEff01}) is usually associated to particle creation, with $|\beta|^2 = \sinh^2r$. However, as we have already said $|\beta|^2$ has to be interpreted here as the {\it number of modes} of a given distribution. When there is no entanglement,  $r \rightarrow 0$, the effective Friedmann equation (\ref{FEff01}) coincides with (\ref{FE21}). The solutions of (\ref{FEff01}) are then essentially the same as those of (\ref{FE21}). However, when the entanglement between the universes is relevant, $\sinh r \gg 1$, the Friedmann equation and therefore the evolution of the universe is significantly modified by the entanglement of the universe with a partner universe.

In the case of the tangential pulse, the entanglement rate is a highly decreasing function of the scale factor so the effect rapidly disappears. However, in the very early stage of the universe the departure from the evolution of a non-entangled universe may be significant. This opens the possibility to detect observational imprints of the multiverse in the properties of the universes for more realistic models. Let us notice that a departure from the exponential expansion of a de Sitter spacetime in the very early stage of the evolution would induce observable effects in the properties of the power spectrum of the CMB \cite{Bouhmadi2011, Scardigli2011, Bouhmadi2013}. Furthermore, an interacting scheme like the one depicted in (\ref{H21}) could modify the processes of vacuum decay in the multiverse \cite{salva2015}, which in turn might induce observable consequences \cite{Bousso2014, Bousso2015}. Other imprints that have been proposed \cite{Laura09,kinney16,Laura16a,Laura16b} should be analyzed in the context of the cyclic multiverse, too.

It is also important to notice that the effects of the inter-universal entanglement are not necessarily  restricted to the very early stage of the universes. We have shown in this paper that in the case of the sinusoidal pulse the entanglement rate is also important when the universes approach the {\it maximum expansion} point. There, the second term in (\ref{FEff01}) becomes dominant and the evolution of the universe turns out to effectively be controlled by the entanglement between the branches of the multiverse. That is an example of observable effects of inter-universal entanglement in an otherwise highly macroscopic and very evolved universe.


\section{Conclusions}
\label{conclusions}

We have studied the possible creation and evolution of parallel cyclic universes evolving within the multiverse which may allow different physical constants and the same geometry. These universes are classically disconnected, but quantum-mechanically entangled and so one is able to apply the thermodynamics of entanglement theory which is known from many physical contexts. We have shown that the entropy of entanglement is large at the big bang and big crunch singularities of the parallel universes as well as at the maxima of the expansion of individual universes. The latter confirms some earlier studies that quantum effects are strong at the turning points of the evolution of the universes (i.e. for macroscopic universes) -- the result was obtained on the base of the formalism of the timeless Wheeler-DeWitt equation and decoherence. Such effects (though related to the same universe) were studied already in quantum cosmology \cite{tunnel95,Mithani12,Mithani14}. In our scenario it requires at least two parallel universes (the ``doubleverse'' of Ref.~\cite{Marosek2016}), for which one can have one universe being replaced quantum-mechanically due to a tunneling effect into the second universe at their maximum expansion points. 

Our studies have also shown that the entropy of entanglement at the big rip singularities goes to zero despite the fact that we deal with apparently Planck density macroscopic universes (which violate the null energy condition) and they should, according to the above statement, be of a quantum nature. However, the vanishing of the entanglement seems to be the property of a big rip singularity which leads to a total dissociation of the universe/multiverse structures into infinitely separated patches which loose any sign of entanglement.

The multiverse that we have studied is quantum-mechanically entangled and there are periods of its evolution where the entanglement matters (here the classical singularities such as the big bang and the big rip as well as maximum expansion points) and can lead to an effect of an exchange of the universes by quantum-mechanical tunneling.  However, the relation between classicality and entanglement still should be sorted out in the context of the quantum multiverse.

In quantum optics, the sharp quantum character of the entangled states comes from the fact that the photon distribution that corresponds to a two-mode entangled state of the electromagnetic field does not satisfy certain classical inequalities \cite{Reid1986}. This violation clearly reveals  that the description of the electromagnetic field in terms of photons as individual and independent entities is not appropriate in the regimes where this violation occurs unless we consider as well non-local interactions among them, irrespective of the distance they are separated, which is a highly non-classical assumption.

On cosmological grounds, it means that the quantum character of the inter-universal entanglement is directly related to the independence of the state of the universes and the presence or the absence of non-local interactions in the minisuperspace. It implies that if we consider the multiverse as the most general scenario in cosmology, which is favored by fundamental theories like the string theories, then, we are forced to consider as well interactions among the universes of the multiverse. In that case, the properties and the evolution of the universe, mainly during the very early phase of its evolution but, as we have shown, as well during other stages like the turning point in the case of cyclic universes, would depend not only on the internal properties of the universe but also on the global properties of the whole multiversal state.

A different question is the quantum nature of the universe in terms of the fluctuations of the matter fields. Let us first notice that  the entangled universes considered in the paper are quantum-mechanically represented by WKB wave functions that are valid for values of the scale factor for which, $S(a) \gg \hbar$. In that case, the fluctuations of the spacetime are largely suppressed, the eigenvalue of the quantum momentum is highly peaked around the classical value and, thus, a time variable can be chosen so that the scale factor satisfies the momentum constraint, which is the Friedmann equation. In that sense, the evolution of the spacetime is classical.

However, we know that quantum fluctuations become dominant not only at the big bang and big crunch singularities but also at the turning point of a cyclic universe \cite{packets} as well as at the big rip singularity \cite{PRD2006}. Then, if the degree of entanglement between the states of the universes is related to the quantumness of their matter fields, then, the entropy of entanglement, which is the standard measure of entanglement, might not be the most reliable measure of quantumness because, at least in the case of the big rip singularity, it goes to zero despite the quantum behavior of the matter field that propagate therein \cite{PRD2006}. It seems that a more reliable indicator of the quantum character of the universes could be the temperature of entanglement, which grows to infinity whenever the state of the universe approaches a classically forbidden region, at least in the cases considered in this paper: big bang, big crunch, turning point, and big rip, ($a\rightarrow0$
  in the first two cases,  $a\rightarrow \frac{1}{\sqrt{\Lambda}}$ in the turning point of the sinusoidal pulse, and $t\rightarrow \frac{\pi}{2\sqrt{\Lambda}}$ in the tangential pulse, see Figs.~\ref{fig6sin}--\ref{fig6tan}).

On the other hand, the results obtained in this paper clearly show that entanglement is directly related to the separability of the quantum states of a given representation. In our case, this is represented by the quantum independence of the opposite modes of the diagonal representation, i.e.~the modes that represent opposite branches from the point of view of internal observers, provided that the multiverse stays in the ground state of an invariant representation, regardless of the semiclassical character of the branches. The representations considered here are the physically relevant in the cosmological problem we are dealing with. However, it is worth noticing that the consideration of different representations, which would ultimately be induced by the consideration of different boundary conditions, could have thrown different rates of entanglement. Thus, entanglement is directly connected with a representation problem, i.e.~what representation has to be chosen to represent the physical system under consideration, and once this is fixed, it is also related to the correlated properties of two classically disconnected (separated) subsystems.

Finally, a separate problem is what one means by the notion of the universe within the framework of the multiverse using, for example, the hierarchy given in Ref.~\cite{TegmarkSA}. If we use the antipodal symmetry for the time variables of the consecutive branches like it is depicted in Fig.~\ref{fig5}, then, all branches are quantum-mechanically exact copies of each other except for the internal processes given in the particular branches, which should be randomly distributed along the finite number of possibilities. Thus, the multiverse depicted in this paper could be interpreted as a Level III multiverse because in an infinite number of universes all probable distributions of the internal degrees of freedom would be accounted for (in fact, an infinite number of times). However, as it is pointed out in Ref.~\cite{Tegmark2004}, this Level III multiverse would represent nothing more than a Level I multiverse, i.e.~an infinite number of Hubble volumes, if the fundamental constants are 
 taken to be the same in all universes, or a Level II multiverse if instead, different values and functions are taken for the fundamental (varying and not varying) constants.

\vspace{0.7cm}

\section*{Acknowledgments}

The work of M.\,P.\,D., A.\,B.~and M.K.~was supported by the Polish National Science Center (NCN) under Grant No.~DEC-2012/06/A/ST2/00395. The work of S.\,R.-P.~was supported partially by the project FIS2012-38816 from the Spanish Ministerio de Econom\'\i a y Competitividad.


\bibliographystyle{apsrev}

\end{document}